\documentclass[reprint,superscriptaddress,amsmath,amssymb, aps, prl, floatfix,]{revtex4-2}

\usepackage{graphicx} 
\usepackage{dcolumn} 
\usepackage{sidecap}
\usepackage{float}
\usepackage{physics}
\usepackage{bm}  

\usepackage[svgnames]{xcolor}
\usepackage[bookmarks=false,linkcolor=blue,urlcolor=blue,colorlinks,citecolor=blue]{hyperref}

\usepackage{lipsum}
\usepackage{booktabs}

\begin{document}

\title{A minimal and universal representation of fermionic wavefunctions
\\ (fermions = bosons + one)
}

\author{Liang Fu}
\affiliation{Department of Physics, Massachusetts Institute of Technology, Cambridge, MA 02139, USA}


\begin{abstract}

Representing fermionic wavefunctions efficiently is a central problem in quantum physics, chemistry and materials science. In this work, we introduce a universal and exact representation of continuous antisymmetric functions by lifting them to continuous symmetric functions defined on an enlarged space. 
Building on this lifting, we obtain a \emph{parity-graded representation} of fermionic wavefunctions, expressed in terms of symmetric feature variables that encode particle configuration and antisymmetric feature variables that encode exchange statistics. This representation is both exact and minimal: the number of required features scales as $D\sim N^d$ ($d$ is spatial dimension) or $D\sim N$ depending on the symmetric feature maps employed.   
Our results provide a rigorous mathematical foundation for efficient representations of fermionic wavefunctions and enable scalable and systematically improvable neural network solvers for many-electron systems. 
\end{abstract}

\maketitle

A fundamental principle of quantum mechanics is that identical particles are indistinguishable. Consequently, exchanging two identical particles can change the many-body wavefunction only by an overall phase factor: $+1$ for bosons (symmetric) and $-1$ for fermions (antisymmetric). The antisymmetry of fermionic wavefunctions embodies the Pauli exclusion principle, which underlies the stability of matter \cite{LiebSeiringer} and governs the electronic properties of metals and insulators \cite{Bloch}.

Despite its central importance in physics, chemistry and materials science, efficiently representing  fermion states remains a major challenge. In essence, the problem is this: how can one efficiently express an arbitrary $N$-fermion wavefunction $\psi(\bm r_1,...,\bm r_N)$, which satisfies the antisymmetric condition   
\begin{equation}
\psi(\mathbf{r}_1, ..., \mathbf{r}_N) = (-1)^\sigma \psi(\mathbf{r}_{\sigma(1)},\ldots,\mathbf{r}_{\sigma(N)}),  \quad \sigma \in S_N \label{antisymmetry}
\end{equation}
in terms of unconstrained functions?  
The standard approach is to expand $\psi$ in terms of Slater determinants constructed from a  single-particle basis set $\phi_1(\bm r), ...,\phi_L(\bm r)$. This representation becomes exact only in the complete basis set limit ($L\rightarrow \infty$). Even when truncated to a finite basis with $L\propto N$, the number of required Slater determinants still scales \emph{exponentially} with particle number $N$.  Alternatively, one may construct $\psi$ by explicitly antisymmetrizing a general function over all particle permutations, but this approach is also inefficient due to the $N!$ permutations involved.  


Historically, a class of fermionic wavefunctions was constructed by multiplying a Slater determinant by a Jastrow factor~\cite{Jastrow} and incorporates backflow transformations of particle coordinates~\cite{FeynmanCohenBackflow, Ceperly}.  
More recently, neural-network-based fermionic wavefunctions~\cite{LuoClarkBackflowNN, PauliNet, MPNQS2024}---most prominently the FermiNet~\cite{FermiNet}---have extended this backflow-determinant framework, achieving greatly enhanced expressive power and impressive accuracy. 

Although a sufficiently large neural network can approximate any \emph{continuous} function over a compact domain arbitrarily well (the universal approximation theorem~\cite{Pinkus}), it remains unclear whether continuous \emph{antisymmetric} functions can be universally represented by existing \emph{fermionic neural networks}—which enforces antisymmetry by design.
Previous arguments~\cite{FermiNet,Hutter} suggesting that a single FermiNet determinant can represent any antisymmetric function rely on sorting the particle coordinates---a procedure that necessarily introduces \emph{discontinuities} in spatial dimensions $d>1$, and is therefore beyond the reach of neural network approximators.    

Recently, in an important and stimulating mathematical work~\cite{ChenLuExactAntiSym}, Chen and Lu showed that any continuous antisymmetric function can be expressed as a composition of an odd function with a fixed set of antisymmetric basis functions---or feature functions in the language of machine learning. In contrast to the Slater determinant basis, the required number of such functions, $D_{a}$, scales only \emph{polynomially} with particle number $D_a \sim N^{d+2}$ for $d>1$ and $N\gg d$. This result is the first efficient representation of fermionic wavefunctions.  

In this work, we introduce a new approach to exactly represent antisymmetric functions and provide a new universal representation. We show that \emph{any} continuous antisymmetric function $\psi(\bm r_1, ..., \bm r_N)$ can be lifted to a continuous function $\Psi(\bm r_1, ...., \bm r_N, \bm \eta)$ by embedding particle coordinates into an enlarged space that includes auxiliary coordinates $\bm \eta$. The lifted function $\Psi$ 
recovers $\psi$ when restricted to the physical sector $\bm \eta=\bm \eta(\bm r_1,...,\bm r_N)$ as specified by an \emph{antisymmetric} mapping from particle coordinates to $\bm \eta$ coordinates, and $\Psi$ is \emph{symmetric} in particle coordinates $\bm r_1,...,\bm r_N$. Equivalently, this construction provides an exact transformation from fermions to bosons through the introduction of just one ancilla particle that carries the ``signature'' of Fermi statistics.      

Building on this lifting, we obtain a universal representation: any continuous antisymmetric function $\psi(\bm r_1,...,\bm r_N)$ can be expressed as a continuous function $f(\bm \xi, \bm \eta)$ of ``generalized coordinates''---namely, \emph{symmetric} feature variables $\bm \xi$ that encode particle configuration in quotient space $(\mathbb{R}^d)^N/S_N$ and \emph{antisymmetric} feature variables $\bm \eta$ that encode Fermi statistics.   
This ``parity-graded'' representation guarantees fermionic antisymmetry without requiring any explicit constraint on $f$. It is both exact and minimal. The feature dimension scales with particle number as $D \sim N^d$ when a standard set of symmetric features is employed. Using a random  selection of symmetric features~\cite{DymGortler2205}, the feature dimension in our representation can be further reduced to $D \sim  N$ for $N\gg d$ in any spatial dimension.

The key component of our representation is the antisymmetric feature function $\bm \eta(\bm r_1, ...., \bm r_N)$---the encoder of Fermi statistics that links $\psi$ and $\Psi$. We identify \emph{canonical} choices of $\bm \eta$ that are independent of the target wavefunction $\psi$ and provide their explicit forms. For $d=2$, $\bm \eta$ can be realized by certain quantum Hall wavefunctions in the lowest Landau level.

\section{Fermion-Boson Transformation}

Our basic idea is simple. Given any continuous  antisymmetric function  $\psi(\mathbf{r}_1, ..., \mathbf{r}_N)$, we construct a lifted function  $\Psi(\mathbf{r}_1,...,\mathbf{r}_N,\bm \eta)$ defined on an enlarged space that includes an auxiliary variable $\bm \eta$. On the ``physical sector'', specified by a mapping $\bm \eta=\bm \eta \bm(\bm r_1, ..., \bm r_N)$, the lifted function reduces to the original one: 
\begin{equation}
\psi(\mathbf{r}_1,...,\mathbf{r}_N) = \Psi\left(\mathbf{r}_1,...,\mathbf{r}_N,\bm \eta(\mathbf{r}_1,...,\mathbf{r}_N) \right).   \label{psi-Psi}
\end{equation} 
We show that for a suitably chosen mapping $\bm \eta$, the lifted function $\Psi$ admits a much simpler representation, from which an efficient and universal representation of antisymmetric functions follows.    

Let $\psi:\Omega \to \mathbb{C}$ be a continuous antisymmetric function of $\bm R=(\mathbf{r}_1,\ldots,\mathbf{r}_N)$,  defined on a compact domain $\Omega \subset (\mathbb{R}^d)^N$. 
Let $\bm\eta:\Omega \to \mathbb{R}^c$ be a feature map that maps each physical coordinate to $c$ auxiliary variables,  
$\bm\eta(\bm R) = (\eta_1(\bm R), \ldots, \eta_c(\bm R))$.
Our construction uses a suitably-chosen feature map that satisfies the following required properties.

\textbf{Property 1.} The feature map is continuous and antisymmetric:  for any $a=1,...,c$ and any permutation $\sigma \in S_N$, 
\begin{equation}
\eta_a(\mathbf{r}_1, ..., \mathbf{r}_N) = (-1)^\sigma \eta_a(\mathbf{r}_{\sigma(1)},\ldots,\mathbf{r}_{\sigma(N)}). 
\end{equation}
In particular, Property~1 implies that $\bm\eta(\bm R)=\bm 0$ (the zero vector in $\mathbb{R}^c$)  whenever $\bm R$ contains a \emph{collision}, i.e., $\mathbf r_i=\mathbf r_j$ for some $i\neq j$.

\textbf{Property 2.} Conversely, $\bm\eta(\bm R)=\bm 0$ \emph{only if} $\bm R$ contains a collision.  

A feature map $\bm \eta$ that satisfies Properties~1 and 2 is able to (1) exactly detect every collision-free configuration; and (2) distinguish even- and odd-permutations of any collision-free configuration through opposite signs of $\bm \eta$.  We refer to such a map as a \emph{signature encoder}.  

Assuming a signature encoder exists, we \emph{define} a lifted function on the union of two \emph{opposite embeddings} in the enlarged space, $(\bm R, \bm \eta_1)$ with $\bm \eta_1 = \bm \eta(\bm R)$ and $(\bm R, \bm \eta_2)$ with $\bm \eta_2 =-\bm \eta(\bm R)$, as follows:  
\begin{eqnarray}
\Psi (\bm R,  \bm \eta_1)  = \psi(\bm R),  
\qquad
\Psi (\bm R, \bm \eta_2) = -\psi(\bm R).  \label{Psi-def}
\end{eqnarray}
Importantly, this definition is well defined. By Property 1, for any collision-free configuration, we have $\bm \eta_1 \neq \bm \eta_2$, so the two values $\Psi(\bm R, \bm \eta_1)$ and $\Psi(\bm R, \bm \eta_2)$ can be assigned separately according to Eq.\eqref{Psi-def}. For any collision configuration, we have $\bm \eta_1=\bm \eta_2=0$, and simultaneously $\psi(\bm R)=0$ due to antisymmetry, which remains consistent with Eq.\eqref{Psi-def}. 
Thus, the lifted function $\Psi$ realizes 
$\psi$ as a two-sheeted cover of configuration space  distinguished by opposite embeddings $\pm \bm \eta(\bm R)$.

By its definition, the lifted function has the following properties:    

\textbf{Property 3.} $\Psi$ is \emph{odd} in the auxiliary variable $\bm \eta$:  
\begin{equation}
\Psi(\mathbf{r}_1,...,\mathbf{r}_N, -\bm \eta ) = - \Psi(\mathbf{r}_1,...,\mathbf{r}_N, \bm \eta ). \label{Psi-odd}
\end{equation} 
and 

\textbf{Property 4.}  $\Psi$ is \emph{symmetric} under particle permutation in $\bm R$---that is, for  any permutation $\sigma$, 
\begin{equation}
\Psi(\mathbf{r}_1,...,\mathbf{r}_N, \bm \eta )= \Psi (\mathbf{r}_{\sigma(1)},...,\mathbf{r}_{\sigma(N)}, \bm \eta ).
\label{Psi-even}
\end{equation}
This follows because both $\psi$ and $\bm\eta$ acquire the same sign factor $(-1)^\sigma$ 
under permutations due to antisymmetry. Since $\Psi$ is odd in $\bm \eta$ (Property 3), these sign changes cancel, 
leaving the lifted function invariant under particle permutations in the enlarged space.

By continuity, the lifted function, initially defined on the two opposite embeddings, 
can be extended to a continuous function on a larger region of the $(\bm R,\bm \eta)$-space (Tietze extension theorem).  
Conversely, for any continuous function $\Psi(\bm R,\bm \eta)$ in the enlarged space that 
satisfies Eqs.~\eqref{Psi-odd} and~\eqref{Psi-even}, restricting to the ``physical sector'' defined by the signature encoder---that is, setting $\bm \eta=\bm \eta(\bm R)$---always yields an antisymmetric function. 

Therefore, by the above procedure employing a signature encoder, any continuous \emph{antisymmetric} function $\psi(\bm R)$ can be lifted to a continuous function $\Psi(\bm R,\bm \eta)$ that is  \emph{invariant} under the permutation of particle coordinates $\bm R$ and \emph{odd} in auxiliary coordinate $\bm \eta$, and can be recovered from it.  Thus, the task of representing $\psi$ can be achieved by finding a universal representation of $\Psi$ in the enlarged space.  
This constitutes our first main result. 

\textbf{Remark.} The lifted function $\Psi(\bm r_1,\ldots,\bm r_N,\eta_1,\ldots,\eta_c)$ can be interpreted as the wavefunction of a system of $N$ bosons in  $d$ dimensions, augmented by a single ancilla  particle in  $c$ dimensions.  The ancilla particle is restricted to odd-parity modes in $\bm \eta$ space and is entangled with the $N$ bosons. Remarkably, any state of $N$ fermions can be retrieved from the corresponding state of this system. In this sense, the bosonic wavefunction $\Psi$ 
captures fermionic antisymmetry of $\psi$ through the addition of just one ancilla particle.  

\section{Signature Encoder}


The signature encoder $\bm \eta(\bm R)$ plays the essential role in recovering the fermionic 
wavefunction $\psi(\bm R)$ from the bosonic wavefunction $\tilde{\Psi}(\bm R,\bm \eta)$; equivalently, it implements the boson-fermion transformation. In this section, we present the explicit form of  $\bm \eta(\bm R)$.   

By Property 1,  $\bm \eta(\bm R)=(\eta_1(\bm R),...,\eta_c(\bm R)) \in \mathbb{R}^c$ may be viewed as a collection of real-valued fermionic functions, each $\eta_a$ changing sign under odd permutations. Property 2 requires that these functions vanish \emph{simultaneously only} on  collision configurations.  
This requirement is nontrivial: by continuity, the zero set of a single real-valued fermionic function typically forms a codimension-1 submanifold of configuration space, whereas the collision set $\{\bm r_i=\bm r_j, i \neq j \}$ has codimension $d$. Thus, for $d>1$, a single fermionic function generally has far more zeros than the collision locus mandated by the Pauli principle. 

The natural question, then,  is whether one can choose \emph{a finite family} $\eta_1,...,\eta_c$ so that at every collision-free configuration at least one component is nonzero---equivalently, the common zero set
$\{\bm \eta=\bm 0\}$ coincides with the collision set---and, if so, what is the \emph{minimal}  dimension $c$ required?   

\subsection{One and Two Dimensions}

We now present a minimal set of antisymmetric real functions that serve as signature encoders in one and two dimensions. 
\begin{eqnarray}
&d=1:& \eta= \prod_{i<j} (x_i - x_j) \label{d=1} \label{d=1}  \\
&d=2:& \eta_1 + \bm i \eta_2 =  \prod_{i<j} (z_i -z_j)  \label{d=2} 
\end{eqnarray}
where $z_i=x_i + \bm i y_i$ are complex coordinates. 

For $d=1$, a single function  $\eta$---the real Vandermonde polynomial of $x_1,..., x_N$---suffices: as a product of all pairwise differences $x_i-x_j$, it is antisymmetric and vanishes \emph{only} on the collision set $\{ x_i=x_j, i \neq j \}$, thus satisfying Property~1 and 2.   

For $d=2$, two antisymmetric real functions $\eta_1,\eta_2$ of the coordinates $(x_1,y_1),...,(x_N, y_N)$ are required. They correspond to the real and imaginary parts of a complex Vandermonde polynomial of the complex coordinates, as given in Eq.~\eqref{d=2}.  
As a product of $z_i-z_j$, the pair $(\eta_{1}, \eta_2)$ vanishes simultaneously  \emph{only} if $z_i=z_j$ or equivalently  $(x_i, y_i)= (x_j, y_j)$ for some $i \neq j$.  

Eqs.\eqref{d=1} and \eqref{d=2} can be generalized into pair product wavefunctions as follows:  
\begin{eqnarray}
&d=1:& \quad  \eta=\prod_{i<j} \Delta(x_i - x_j)\label{d=1}  \\
&d=2:&  \quad \eta_1 + \bm i \eta_2= \prod_{i<j} \Delta(\bm r_i - \bm r_j),  
\end{eqnarray}
with $\Delta$ any odd function that vanishes \emph{only} at the origin. As a further generalization, $\eta$ may be multiplied by a symmetric product factor $\prod_i \rho(\bm r_i)$ with $\rho(\bm r)$ any function that is everywhere nonzero.  

As an interesting example in $d=2$, a signature encoder is realized by  quantum Hall wavefunctions in the lowest Landau level,
\begin{equation}
\eta_1 + \bm i \eta_2= \prod_{i<j} (z_i - z_j)^q \times \prod_i \exp(-|z_i|^2), 
\end{equation}
where $q$ is an odd positive integer. 
In particular, $q=1$ corresponds to the $\nu=1$ quantum Hall state---a completely full Landau level---whose wavefunction equals a Slater determinant $\det[\phi_i(z_j)]]$ with single-particle orbitals $\phi_i(z)=z^{i-1} \exp(-|z|^2) $. On the other hand, $q=3$ corresponds to the $\nu=\frac{1}{3}$ Laughlin wavefunction, which cannot be written as a single Slater determinant. Both serve as valid signature encoders capable of transforming \emph{any} fermionic wavefunction $\psi(\bm r_1,...,\bm r_N)$ into a corresponding bosonic wavefunction $\Psi(\bm r_1,...,\bm r_N, \bm \eta)$, regardless of whether $\psi$ resides within the lowest Landau level.




\subsection{All Dimensions}

To construct $\bm \eta$ functions in higher dimensions $d>2$, it is tempting to consider generalizations of Vandermonde determinants from the real or complex coordinate to the quaternion coordinate, such as $q=x+\bm iy+\bm j z + \bm k w$ in $d=4$. Because the quaternion ring has no zero divisor, the quaternion Vandermonde polynomial $\prod_{i<j} (q_i-q_j)$ vanishes \emph{only} at collision, thereby satisfying Property 2. However, since quaternions are noncommutative, the quaternion Vandermonde polynomial is generally {\it not} antisymmetric under particle permutation, and thus fails to satisfy Property~1.        

To proceed, we consider a slight variant of the problem---representing antisymmetric wavefunctions subject to \emph{periodic boundary condition}, which describe fermions on a torus $T^d$ rather than on a compact domain in $\mathbb{R}^d$ as considered so far. Mathematically, $\psi$ satisfies      
\begin{equation}
 \psi(...,\bm r_i,...) = \psi (...,\bm r_i + \bm L,...)\
\end{equation}
for any $\bm L=\sum_{l=1}^d n_l \bm L_l$ with $(n_1, ..., n_d) \in \mathbb{Z}^d$, 
where $\bm L_1,..., \bm L_d$ form a set of linearly independent supercell vectors. 

Such a periodic function can be expressed in terms of its Fourier expansion, 
\begin{equation}
 \psi( \bm r_1 ,...,\bm r_N) =\sum_{\bm k_1,...,\bm k_N} e^{i \bm k_i\cdot \bm r_i } \tilde{\psi} (\bm k_1,..., \bm k_N)\
\end{equation}
where the allowed wavevectors $\bm k_i$ satisfy $e^{i \bm k_i \cdot \bm L_l}=1$ for all $i, l$ and thus form a lattice in reciprocal space. Specifically, each wavector can be written as 
\begin{equation}
\bm k_i=\sum_{l=1}^d m_l \bm g_l, \quad (m_1,...,m_d)\in \mathbb{Z}^d 
\end{equation}
where $\bm g_1,...,\bm g_d$ are a complete set of reciprocal supercell vectors defined by $\bm g_k \cdot L_{l} = 2\pi \delta_{kl}$, $k,l=1,...,d$. 

Complementary to $\psi(\bm r_1,...,\bm r_N)$ in \emph{continuous} real space, $\tilde{\psi}(\bm k_1,..., \bm k_N)$ represents the probability amplitude for finding $N$ fermions occupying the set of plane wave modes at \emph{discrete} wavevectors $\bm k_1,...,\bm k_N$. Related through the Fourier transform, $\psi$ and $\tilde{\psi}$ describe the same quantum state of $N$ fermions in first quantization, and both must be antisymmetric under particle exchange. 

The fermion-boson transformation introduced in Section 1 applies equally well to $\tilde{\psi}$.   
We now introduce a single function $\eta(\bm K)$ as a signature encoder for antisymmetric functions $\tilde{\psi}(\bm K)$, where $\bm K=(\bm k_1,...,\bm k_N) \in (\mathbb{Z}^d)^N$, valid in arbitrary dimension.  
\begin{equation}
 \eta(\bm K) =\prod_{i<j} \delta(\bm k_i - \bm k_j)
\end{equation}
with $\delta(\bm k)$ any odd function over the reciprocal lattice that vanishes \emph{only} at the origin. 
A simple choice is 
$\delta(\bm k) = \bm w \cdot \bm k$,   
where $\bm w$ is a fixed vector defining an \emph{incommensurate} direction in reciprocal space, i.e., $\bm w\cdot \bm k=0$ implies $\bm k=0$ for $\bm k$ in reciprocal lattice. Equivalently, the hyperplane $\bm w \cdot \bm k=0$ intersects the reciprocal lattice only at the origin, ensuring that 
$\delta(\bm k)=0$ if and only if $\bm k=\bm 0$. Such incommensurate directions are generic: for almost any choice of $\bm w$, this condition is satisfied.
Hence, the function $\eta(\bm K)$ satisfies Properties~1 and 2 and thus serves as a valid signature encoder. 

To summarize this section, we provide the explicit forms of the signature encoder $\bm \eta$ for representing fermionic wavefunctions in coordinate space, defined either on a compact domain or under periodic boundary conditions. For a compact domain, a convenient choice of signature encoder is Vandermonde polynomial of the real coordinates for $d=1$, or of the complex coordinates for $d=2$. For periodic boundary conditions, a canonical signature encoder can be defined for the dual wavefunction over the reciprocal lattice, given by the Vandermonde polynomial of the one-dimensional projection of the reciprocal lattice coordinates, valid in arbitrary dimension. 

Together with Section 1, this completes the exact transformation from fermions to bosons through the introduction of a single ancilla particle in the $\bm \eta$ space, achieved by Eq.\eqref{psi-Psi},  with the mapping from physical space to $\bm \eta$ coordinate defined by the signature encoders presented above. As demonstrated in all the cases, the dimensionality of the $\bm \eta$ space is minimal: $c=1$ or $2$.

\section{Universal Representation}
Finally, we turn to the universal representation of lifted functions $\Psi$ on $(\bm R,\bm\eta)$-space that satisfy Properties~3 and~4---namely, permutation invariance in $\bm r_1,...,\bm r_N$ and oddness in $\bm \eta$.  The same approach also applies to lifted functions on $(\bm K, \bm \eta)$-space.

Observe that any such $\Psi$ can be
expressed as an odd superposition
\begin{equation}
\Psi(\bm R,\bm \eta)
= \frac{\Psi_0(\bm R,\bm \eta)-\Psi_0(\bm R,-\bm \eta)}{2},
\label{eq:tPsi}
\end{equation}
where $\Psi_0$ is invariant under permutations in $\bm r_1,...,\bm r_N$ and unconstrained in its dependence on $\bm \eta$.

Thus, the task reduces to representing $\Psi_0(\bm R,\bm \eta)$---a family of continuous symmetric functions parameterized by $\bm \eta$. This can be achieved naturally by generalizing the representation of symmetric functions. Following the \emph{Deep Sets} framework \cite{DeepSets}, universal representation of symmetric functions has been extensively studied, and efficient constructions are now well established \cite{MaronEtAl2019, WangYangLiWangLiICLR2024, ChenChenLu2024}.   

The key idea is that any continuous symmetric function $\phi(\bm R)$ can be represented by the composition of a \emph{permutation-invariant} continuous map from particle coordinate $\bm R$ to a $m$-dimensional feature vector $\bm \xi=(\xi_1,...,\xi_m) \in \mathbb{R}^m$, followed by a continuous function $f$ defined on the feature space:  
\begin{equation}
\phi(\bm R) = f(\bm \xi(\bm R)). \label{phi-rho}
\end{equation}
Importantly, there exists a feature map  $\bm \xi$ that \emph{uniquely} encodes particle configuration up to permutation, that is, $\bm \xi(\bm r_1,...,\bm r_N)=\bm \xi(\bm r'_1,...,\bm r'_N)$ \emph{if and only if} there exists a  permutation $\sigma$ such that $\bm (\bm r'_1,...,\bm r'_N)=(\bm r_{\sigma(1)},...,\bm r_{\sigma(N)})$. We refer to such a mapping $\bm \xi$ as a \emph{set embedding}, since it provides an injective embedding of the \emph{unordered} set of particle coordinates $\{\bm r_i\}$ (i.e., the configuration space $(\mathbb{R}^d)^N/S_N$) into the feature space. Consequently, any $\phi(\bm R)$ can be expressed as a continuous function of the feature vector, defined as $f(\bm \xi) = \phi(\bm R=\bm \xi^{-1})$. 

Specifically, for symmetric functions on a compact domain, a set embedding can be realized through a set of \emph{multisymmetric power sums} (which are manifestly permutation invariant): 
\begin{equation}
\sum_{i=1}^N r^{k_1}_{i,1}...r^{k_d}_{i,d}, \quad 1 \leq k_1 + ... +k_d \leq N.
\end{equation}
In physics terms, these power sums correspond to multipole moments of particles. 
It is know that the coordinates $\bm r_1, ..., \bm r_N$ can be uniquely recovered up to permutation from the values of these power sums \cite{WeylClassicalGroups, Dalbec1999, Vaccarino2005}. The number of these power sums, which determines the dimensionality of the symmetric feature space  $m$, is $m=\binom{N+d}{d}-1$, which scales as $m\sim N^d$ for $N\gg d$. 
Alternatively, a recent work \cite{DymGortler2205} shows that permutation-invariant embeddings can be achieved by randomly selecting $2dN+1$ symmetric functions, thus reducing the feature space dimension to $O(N)$ for any $d$.

For our problem, $\Psi_0(\bm R,\bm \eta)$ is a symmetric function of $\bm R$ parameterized by $\bm \eta$.  Thus, it can be represented by generalizing Eq.\eqref{phi-rho} to make $f$ dependent on $\bm \eta$. 
Combining the result with Eqs.~\eqref{psi-Psi} and \eqref{eq:tPsi},  
we establish that any continuous antisymmetric function $\psi$ can be exactly represented---through the signature encoder $\bm \eta$ and the set embedding $\bm \xi$---as a continuous function $f$ of the feature space of $(\bm \xi, \bm \eta) \in \mathbb{R}^{m+c}$,
\begin{eqnarray}
 \psi(\bm R) = \frac{f \left(\bm \xi(\bm R) ,\bm \eta(\bm R)\right)-f\left(\bm \xi(\bm R) ,-\bm \eta(\bm R) \right)}{2},
\label{psi-f}
\end{eqnarray}

This representation is universal. It guarantees the antisymmetry of $\psi$ without imposing any explicit constraint on $f$, but rather through the combination effect of (1) the permutation-invariant set embedding $\bm \xi$, (2) the antisymmetric signature encoder $\bm \eta$, and (3) oddness in $\bm \eta$ introduced by antisymmetrization.
Since this representation uses both symmetric ($\bm \xi$) and antisymmetric ($\bm \eta$) features of particle coordinates $\bm R$, we refer to it as a ``parity-graded representation'' of fermionic wavefunctions. This constitutes our second main result. 

Since  our construction only uses a minimal number ($c=1,2$) of antisymmetric features, the feature space dimension in our $(\bm \xi,\bm \eta)$-representation, $D=m+c$, scales with the particle number $N$ in the same way as the feature space scaling for symmetric functions---namely, $D\sim N^d$ or $D \sim N$ depending on the set embedding method employed. For comparison, the universal representation of Ref.\cite{ChenLuExactAntiSym} only employs antisymmetric features, and thus requires a significantly larger feature space dimension $D_a \sim N^{d+2}$ in $d>1$. Hence, our  parity-graded representation constitutes the most efficient method for representing general antisymmetric functions.   

As an example illustrating how the $(\bm \xi, \bm \eta)$-representation works, let us consider representing fractional quantum Hall wavefunctions such as the $\nu=1/3$ Laughlin wavefunction $\psi_{\frac{1}{3}}=\prod_{i<j} (z_i - z_j)^3 \exp(-\sum_i |z_i|^2/2)$, using the signature map $\eta_1 +i \eta_2=\prod_{i<j} (z_i -z_j) \exp(-\sum_i |z_i|^2/2)$. One can write $\psi_{\frac{1}{3}}=(\eta_1 + i \eta_2 )\times P$, where $P=\prod_{i<j} (z_i-z_j)^{2}$ is a symmetric polynomial of complex coordinates $z_1, ..., z_N$. Remarkably, $P$ can be neatly expressed in terms of power sums $\xi_k=\sum_i z_i^k$ through Hankel determinant: 
\begin{equation}
P = \det [\xi_{i+j-2}]_{1\leq i,j\leq N}. 
\end{equation}
Therefore, the $(\bm \xi, \bm \eta)$-representation of $\nu=1/3$ Laughlin wavefunction  
takes an analytical form: 
\begin{equation}
f(\bm \xi, \bm \eta)= (\eta_1 + i \eta_2) \det[\xi_{i+j-2}],        
\end{equation}
using a total of $2N$ feature variables (discounting $\xi_0=N$): $\xi_1, ..., \xi_{2N-2}, \eta_1, \eta_2$.

\section{Discussion}

We emphasize that the function $f$ representing a continuous antisymmetric function $\psi$ is itself guaranteed to be continuous. This follows from the continuity-preserving nature of each step in the construction, $\psi \rightarrow \Psi \rightarrow \Psi_0 \rightarrow f$, and from the use of continuous feature maps $\bm \xi$ and $\bm \eta$.

One may wonder whether the universal representation Eq.\eqref{psi-f} could be further simplified. A seemingly plausible approach is to express $\psi(\bm R)$ as a product of a signature map $\eta(\bm R)$ and a symmetric function $\phi(\bm R)$, where $\phi(\bm R)$ is defined as the ratio $\psi(\bm R)/\eta(\bm R)$. Since $\eta(\bm R)$ is nonzero away from collisions, $\phi(\bm R)$ appears well behaved in that region. However, there is no guarantee that $\phi(\bm R)$ remains continuous in the vicinity of collisions. Consider this example in $d=2$: $\psi=\bar{z}_1-\bar{z}_2$ and $\eta=z_1-z_2$; the ratio $\psi/\eta$ depends on the direction in which $z_1-z_2$ approaches $0$, and thus fails to be continuous near $z_1 = z_2$.         

While  Eq.~\eqref{psi-f} provides an exact and universal representation of continuous antisymmetric functions, the continuous function on feature space $f(\bm \xi, \bm \eta)$ may not inherit the same regularity as the target function $\psi(\bm R)$ in coordinate space. To illustrate this point, consider representing an antisymmetric function  $\psi(x_1, x_2)=x_1-x_2$ using the signature encoder $\eta=(x_1-x_2)^3$ and the set embedding $\bm \xi=(x_1+x_2, x_1^2+x_2^2)$  which uniquely determines the unordered pair $\{ x_1, x_2\}$. Let $E$ denote the image of the feature map $(x_1, x_2) \rightarrow (\xi_1,\xi_2, \eta)$ and define  a region $\Sigma$ of feature space surrounding $E$  defined by $|\eta| \leq 2 D^{3/2}$ and $ D\geq 0$, where $D=2\xi_2-\xi_1^2$. A  function $f$ over $\Sigma$ can then be defined as $f(\xi_1, \xi_2, \eta) = \eta/D$ for $D> 0$ and $0$ on $D=0$. 
Importantly, $f$ defined in this way recovers $\psi$ when restricted to $E$, and is continuous throughout $\Sigma$. Although it provides an exact representation of $\psi$, $f$ is not differentiable at points with $D=0$.  

On the other hand, the universal approximation theorem guarantees that any continuous function $f$---differentiable or not---can be approximated to arbitrary accuracy by a sufficiently large neural network. In the above example, the function $f=\eta/D$ is well approximated by $f_\epsilon = \eta D/(D^2+\epsilon)$ with a small $\epsilon>0$, which is smooth everywhere in $\mathbb{R}^3$ and can be efficiently learned by a neural network. 

Therefore, the $(\bm \xi,\bm \eta)$-representation provides a natural foundation for constructing \emph{universal approximators} for continuous antisymmetric functions through neural networks. In this framework, the antisymmetry of the wavefunction is enforced \emph{exactly} through the parity-graded structure of Eq.~\eqref{psi-f}, while the continuous function $f$ can be implemented as a neural network without any symmetry constraints. 

The set embedding $\xi(\bm R)$, which maps particle coordinates to permutation-invariant features, need not be restricted to analytic forms such as multisymmetric power sums. In practice, $\bm \xi$ can be implemented as a learnable, permutation-invariant architecture---such as the Deep Sets network~\cite{DeepSets} or its extensions~\cite{SetTransformer, MaronEtAl2019, WangYangLiWangLiICLR2024}.
Similarly, the continuous function $f(\bm \xi,\bm \eta)$ may be realized by a deep neural network (for example, a multilayer perceptron) capable of capturing complex correlations between the symmetric ($\bm \xi$) and antisymmetric ($\bm \eta$) features.

This construction offers a principled approach to fermionic neural network wavefunctions in which the antisymmetry is enforced by design, and the \emph{universal representational power} is limited only by the capacity of the neural architectures used for $f$ and $\bm \xi$. In this sense, our 
$(\bm \xi, \bm \eta)$-representation unifies a rigorous mathematical foundation with the flexibility of modern deep learning and opens a direct path to efficient, scalable and systematically improvable neural network solver for Fermi systems.

Finally, we note that the $(\bm \xi, \bm \eta)$ representation naturally incorporates the indistinguishability of identical particles through the definition of the configuration space, which is \emph{not} the Cartesian product of single particle spaces but rather its quotient space $(\mathbb{R}^d)^N/S_N$ \cite{LeinaasMyrheim}. This configuration space is faithfully captured by the symmetric $\bm \xi$ coordinates, while its double covering, represented by the antisymmetric $\bm \eta$ coordinate, gives rise to Fermi statistics.

\newpage


{\it Acknowledgements---}  It is my pleasure to thank Ziang Chen for helpful and informative discussions. I also acknowledge the support from Simons Investigator Award.

\bibliographystyle{unsrt}
\bibliography{refs}

\begin{thebibliography}{10}

\bibitem{LiebSeiringer}
E.~H. Lieb and R.~Seiringer.
\newblock {\em The Stability of Matter in Quantum Mechanics}.
\newblock Cambridge University Press, Cambridge, 2010.

\bibitem{Bloch}
F.~Bloch.
\newblock Über die quantenmechanik der elektronen in kristallgittern.
\newblock {\em Z. Phys.}, 52:555--600, 1929.

\bibitem{Jastrow}
R.~Jastrow.
\newblock Many-body problem with strong forces.
\newblock {\em Phys. Rev.}, 98:1479--1484, 1955.

\bibitem{FeynmanCohenBackflow}
R.~P. Feynman and M.~Cohen.
\newblock Energy spectrum of the excitations in liquid helium.
\newblock {\em Phys. Rev.}, 102:1189--1204, 1956.

\bibitem{Ceperly}
Y.~Kwon, D.~M. Ceperley, and R.~M. Martin.
\newblock Effects of three-body and backflow correlations in the
  two-dimensional electron gas.
\newblock {\em Phys. Rev. B}, 48:12037--12046, 1993.

\bibitem{LuoClarkBackflowNN}
D.~Luo and B.~K. Clark.
\newblock Backflow transformations via neural networks for quantum many-body
  wave functions.
\newblock {\em Phys. Rev. Lett.}, 122:226401, 2019.

\bibitem{PauliNet}
J.~Hermann, Z.~Sch{\"a}tzle, and F.~No{\'e}.
\newblock Deep-neural-network solution of the electronic schr{\"o}dinger
  equation.
\newblock {\em Nat. Chem.}, 12:891--897, 2020.

\bibitem{MPNQS2024}
G.~Pescia, J.~Nys, J.~Kim, A.~Lovato, and G.~Carleo.
\newblock Message-passing neural quantum states for the homogeneous electron
  gas.
\newblock {\em Phys. Rev. B}, 110:035108, 2024.

\bibitem{FermiNet}
D.~Pfau, J.~S. Spencer, A.~G. de~G.~Matthews, and W.~M.~C. Foulkes.
\newblock Ab initio solution of the many-electron schr{\"o}dinger equation with
  deep neural networks.
\newblock {\em Phys. Rev. Research}, 2:033429, 2020.

\bibitem{Pinkus}
A.~Pinkus.
\newblock Approximation theory of the mlp model in neural networks.
\newblock {\em Acta Numerica}, 8:143--195, 1999.

\bibitem{Hutter}
A.~Hutter.
\newblock On representing (anti)symmetric functions.
\newblock {\em arXiv preprint arXiv:2007.15298}, 2020.

\bibitem{ChenLuExactAntiSym}
Z.~Chen and J.~Lu.
\newblock Exact and efficient representation of totally anti-symmetric
  functions.
\newblock {\em arXiv preprint arXiv:2311.05064}, 2023.
\newblock revised Jan 2025.

\bibitem{DymGortler2205}
Nadav Dym and Steven~J. Gortler.
\newblock Low dimensional invariant embeddings for universal geometric
  learning.
\newblock {\em arXiv preprint arXiv:2205.02956}, 2022.

\bibitem{DeepSets}
M.~Zaheer, S.~Kottur, S.~Ravanbakhsh, B.~P{\'o}czos, R.~Salakhutdinov, and
  A.~Smola.
\newblock Deep sets.
\newblock In {\em Advances in Neural Information Processing Systems (NeurIPS)},
  2017.

\bibitem{MaronEtAl2019}
H.~Maron, H.~Ben-Hamu, N.~Shamir, and Y.~Lipman.
\newblock On the universality of invariant networks.
\newblock In {\em Proceedings of the 36th International Conference on Machine
  Learning (ICML)}, volume~97, pages 4363--4371, 2019.

\bibitem{WangYangLiWangLiICLR2024}
P.~Wang, S.~Yang, S.~Li, Z.~Wang, and P.~Li.
\newblock Polynomial width is sufficient for set representation with
  high-dimensional features.
\newblock In {\em Proceedings of the Twelfth International Conference on
  Learning Representations (ICLR)}, 2024.

\bibitem{ChenChenLu2024}
C.~Chen, Z.~Chen, and J.~Lu.
\newblock Representation theorem for multivariable totally symmetric functions.
\newblock {\em Communications in Mathematical Sciences}, 22(5):1195--1201,
  2024.

\bibitem{WeylClassicalGroups}
H.~Weyl.
\newblock {\em The Classical Groups: Their Invariants and Representations}.
\newblock Princeton University Press, Princeton, 2 edition, 1946.

\bibitem{Dalbec1999}
J.~Dalbec.
\newblock Multisymmetric functions.
\newblock {\em Beiträge zur Algebra und Geometrie}, 40(1):27--51, 1999.

\bibitem{Vaccarino2005}
F.~Vaccarino.
\newblock The ring of multisymmetric functions.
\newblock {\em Annales de l'Institut Fourier}, 55(3):717--731, 2005.

\bibitem{SetTransformer}
J.~Lee, Y.~Lee, J.~Kim, A.~Kosiorek, S.~Choi, and Y.~W. Teh.
\newblock Set transformer: A framework for attention-based
  permutation-invariant neural networks.
\newblock In {\em International Conference on Machine Learning (ICML)}, 2019.

\bibitem{LeinaasMyrheim}
J.~M. Leinaas and J.~Myrheim.
\newblock On the theory of identical particles.
\newblock {\em Il Nuovo Cimento B}, 37:1--23, 1977.

\end{thebibliography}

\end{document}